\documentclass[10pt,a4paper]{article}
\pdfoutput=1
\usepackage[margin=1.5in]{geometry}

\usepackage[utf8]{inputenc}
\usepackage{physics}
\usepackage{amsmath}
\usepackage{amssymb}
\usepackage{graphicx}
\usepackage[margin=0.1in]{subcaption}
\usepackage{bbold}
\usepackage{siunitx}
\usepackage{hyperref}
\usepackage{cleveref}
\usepackage{floatrow}
\usepackage[affil-it]{authblk}
\usepackage[
    style=numeric-comp,sorting=none,
    maxcitenames=4, mincitenames=4, 
    maxbibnames=5, minbibnames=5,
    sortcites=true,doi=true,url=false,
    giveninits=true,hyperref]{biblatex}
\bibliography{biblio.bib}

\newcommand{\brho}{\boldsymbol{\rho}}

\newcommand{\F}[1]{\mathcal{F}\left\{ #1 \right\}}

\usepackage[normalem]{ulem} 

\begin{document}
	\title{\textbf{Källén–Lehmann Spectral Representation of the Scalar SU(2) Glueball}}

	\author{David Dudal$^{a, b}$\thanks{\href{mailto:david.dudal@kuleuven.be}{david.dudal@kuleuven.be}}, Orlando Oliveira$^c$\thanks{\href{mailto:orlando@uc.pt}{orlando@uc.pt}}, Martin Roelfs$^{a}$\thanks{\href{mailto:martin.roelfs@kuleuven.be}{martin.roelfs@kuleuven.be}; Corresponding author}}
	\affil{\footnotesize $^{a}$ KU Leuven Campus Kortrijk--Kulak, Department of Physics, Etienne Sabbelaan 53 bus 7657, 8500 Kortrijk, Belgium\\
		$^{b}$ Ghent University, Department of Physics and Astronomy, Krijgslaan 281-S9, 9000 Gent, Belgium\\
		$^{c}$ CFisUC, Department of Physics, University of Coimbra, 3004-516 Coimbra, Portugal}
	\date{}

	\maketitle
	\begin{abstract}
		The estimation of the  K\"all\'en-Lehmann spectral density from gauge invariant lattice QCD two point correlation functions is proposed, and explored via an appropriate inversion method.
		As proof of concept the SU(2) glueball spectrum for the quantum numbers $J^{PC} = 0^{++}$
		is investigated for various values of the lattice spacing. The spectral density and the glueball spectrum are estimated using the published data of \cite{Yamanaka:2019yek}.
		Our estimates for the ground state mass are in good agreement with the traditional approach published therein, which is based on the large time exponential behaviour of the correlation functions.
		Furthermore, the spectral density also contains hints of excites states in the spectrum. 
	\end{abstract}
	
	\section{Introduction}
	
	A major effort to understand the dynamics of strong interactions is the computation of the hadronic spectra.
	In QCD, hadrons, namely mesons and baryons, are seen as bound states of quarks and gluons. Besides the conventional hadronic
	states, multiple-quark and pure glue bound states are also predicted by theory. Of these exotic states, so far only multi-quark states have
	been identified experimentally \cite{Zyla:2020zbs}.  For glueballs there are a number of candidate states \cite{Teper:1998kw,Morningstar:1999rf,Chen:2005mg}, but no established observation.
	Glueballs continue to attract a great deal of theoretical \cite{Narison:1996fm,Brower:2000rp,Ishii:2002ww,Szczepaniak:2003mr,Dudal:2010cd,Sanchis-Alepuz:2015hma,Huber:2020ngt} and experimental attention, with an ongoing effort to identify these pure glue states unambiguously  \cite{Mathieu:2008me,Crede:2008vw,Ochs:2013gi,Sarantsev:2021ein}. Additionally, glueball states have also been discussed as possible candidates for dark matter \cite{Yamanaka:2019yek,Hertzberg:2020xdn}.
	
	A first principles approach to access the hadron spectra, which has been advancing since the beginning of the eighties from last century \cite{Ishikawa:1982tb},
	is lattice QCD. In a typical lattice QCD computation of the bound state mass, an appropriate two point correlation function is evaluated and, from its large time decay behaviour and the associated slope, the ground state mass is estimated \cite{Montvay:1994cy}. In the discrete setting, it is in general tricky to access a priori continuum quantities from finite volume data, \cite{Maiani:1990ca}, although ways to extract more than naively expected from finite volume input have been devised \cite{Lellouch:2000pv}.
	It is common practise in the computation of bound masses to rely on techniques that improve the signal to noise ratio of the
	Monte Carlo simulation. These techniques allow access not only to the ground state mass with the chosen quantum numbers, but, sometimes, also allow the first excited state to be extracted. In practice however, the computation of the excited states masses has proven to be a difficult task, usually based on a generalized eigenvalue problem, as in \cite{Morningstar:1999rf,Chen:2005mg}.
	
	Herein, we aim to discuss an alternative way of accessing the particle masses: via the computation of the K\"all\'en-Lehmann spectral
	representation associated with the momentum space particle propagator \cite{Peskin:1995ev}.
	A possible advantage of using spectral representations compared to a conventional lattice calculation is that it does not necessarily require the use of smearing, or other techniques, to improve the Monte Carlo signal to noise ratio, see e.g.
	\cite{Blossier:2009kd,Athenodorou:2020ani} and references therein.
	As seen below, the computation of the spectral function allows simultaneous extraction of the ground state and, to some extent, also of the 1st excited state, at least for good enough data.
	
	The interest in the K\"all\'en-Lehmann spectral representation goes beyond the determination of particle spectra. Besides
	accessing the spectra at zero and finite temperature, the spectral representation is linked with the analytical structure of the
	associated propagator \cite{Negele:1988vy}, and it allows the computation of thermodynamical and transport properties \cite{Meyer:2011gj}.
	Furthermore, in QCD, or any other confining theory, it can help to understand the confinement
	mechanism \cite{Dudal:2010wn,Dudal:2013yva,Solis:2019fzm,Dudal:2019aew,Binosi:2019ecz,Siringo:2016jrc,Cyrol:2018xeq}. In particular for the glueball sector, future work should focus on the thermal widening of the spectral peaks, indicative of a melting pattern of the glueball modes across (or at least near to) the deconfinement transition. Such information cannot be directly traced from the aforementioned large time slope and requires a more in depth study of the spectral functions themselves \cite{Ishii:2002ww}. A first step set in this paper is developing an efficient inversion scheme at zero temperature that can be quite easily generalized to finite temperature afterwards.
	
	In the current work we focus on the SU(2) glueball states with quantum numbers  $J^{PC} = 0^{++}$,  i.e.~the scalar glueball. However, in principle,
	the procedure can be extended to other quantum numbers and other gauge groups. As discussed below, the masses of the glueball
	$0^{++}$ obtained from the spectral function are in good agreement with the estimates of a more conventional mass calculation.
	
	Let us describe our procedure to access the K\"all\'en-Lehmann spectral representation $\rho(\omega)$ from a two point correlation function $G(p^2)$.
	The relation between these two functions reads
	
	\begin{equation}
		G(p^2) = \int_{0}^{\infty} \frac{2 \omega \rho(\omega) \dd{\omega}}{\omega^2 + p^2} \label{eq:propagator_gl}
		= \int_{-\infty}^{\infty} \frac{\rho(\omega) \dd{\omega}}{\omega - i p}  \ .\notag
	\end{equation}
	This definition assumes that the integrations are well defined \cite{Weinberg:1995mt} and, therefore, $\rho ( \omega)$ approaches
	zero sufficiently fast as $| \omega | \rightarrow \infty$.
	Certain correlators contain polynomial terms that diverge for large $p$ as happens e.g.~for the glueball operator, something which follows immediately from power counting and a dimensional analysis.  However in these cases it is still possible to write down a sensible K\"all\'en–Lehmann spectral representation, if one first subtracts the polynomial part \cite{Weinberg:1995mt}. This corresponds to adding appropriate contact counterterms. So, if $G(p^2)$ has a polynomial part, the propagator can be written as
	\begin{align}
		G(p^2)
		&= \sum_{k=0}^{n-1} a_{k} (p^2-\bar{p}^2)^{k} +  (- p^2 + \bar{p}^2)^{n} \int_{0}^{\infty} \frac{2 \omega \tilde{\rho}(\omega) \dd{\omega}}{\omega^2 + p^2} \label{eq:expanded_propagator}\\
		&=  \sum_{k=0}^{n-1} a_{k} (p^2-\bar{p}^2)^{k} +  (- p^2 + \bar{p}^2)^{n} \int_{-\infty}^{\infty} \frac{\tilde{\rho}(\omega) \dd{\omega}}{\omega - ip} \notag
	\end{align}
	with
	\begin{align}
		a_{k} &= \frac{1}{k!} \pdv[k]{G(p^2)}{(p^2)}\Bigg\vert_{p^2=\bar{p}^2}, \label{eq:taylor_coeff} \\
		\tilde{\rho}(\omega) &= \frac{\rho(\omega)}{(\omega^2 + \bar{p}^2)^n},
	\end{align}
	and $\bar{p}^2$ is a reference momentum scale at which the subtraction is done.
	A derivation of the above relations can be
	found in \cref{app:subtractions}.
	For the scalar glueball a dimensional analysis \cite{Dudal:2010cd} shows that $n=3$. The spectral function
	$\rho(\omega)$ can thus be obtained, if the subtraction of the polynomial part can be performed.
	However, performing subtractions on numerical data is extremely sensitive to the choice of $\bar{p}$ and results in relative rapid
	variation of $\{ a_k \}$ with the subtraction point \cite{Oliveira:2012eu}.
	
	An elegant way to perform the subtractions, without actually having to do these, is by considering the Fourier transform of $G(p^2)$ and look at the Schwinger function defined as
	\begin{equation}
		C(\tau) = \F{G(p^2)}(\tau) = \int_{-\infty}^{\infty} G(p^2) \bigg\vert_{\vec{p} = 0, p_4 \neq 0} ~ e^{-i p_4 \tau} \dd{p_4}.
		\label{eq:Schwinger}
	\end{equation}
	Setting the subtraction reference momentum $\bar{p} = 0$, yields
	\begin{align}
		C(\tau) &= \F{G(p^2)}(\tau) \\
		&= \mathcal{L}\left\{ \rho(\omega) \right\}(|\tau|) + 2 \pi \sum_{k=0}^{n-1} a_{k} (-1)^k \delta^{(2k)}(\tau) \\
		&\quad + 4 \pi (-1)^{n+1} \sum_{k=2}^{n} \delta^{(2(n-k)+2)}(\tau) \int_{0}^{\infty} \dd{\omega} \omega^{2k-3} \tilde{\rho}(\omega)\, , \notag
	\end{align}
	see \cref{app:subtractions} for more details.
	The important observation however, is that $C(\tau) = \mathcal{L}\left\{ \rho(\omega) \right\}(|\tau|)$ when $\tau~\neq~0$, and equal to a sum of (derivatives of) Dirac delta functions when $\tau = 0$.
	Therefore, $\rho(\omega)$ can be recovered by taking $C(\tau)$ for $\tau>0$, and inverting the Laplace transformation.
	
	Because the inverse Laplace transform is an ill-defined numerical problem, regularization is necessary in order to perform the
	inversion. We shall use Tikhonov regularization, similar to our previously published method~\cite{Dudal:2013yva,Dudal:2019gvn}. However,
	because glueballs are observable particles, their spectral density $\rho(\omega)$ is non-negative. 
	Therefore we shall implement Tikhonov
	regularization using Non-Negative Least Squares (NNLS) \cite{Lawson1995}, to ensure
	a positive spectral function $\rho(\omega) \geq 0$. Tikhonov regularization is not the only possible way to regularize the inversion,
	and a number of different regularization strategies have been explored in literature to access the spectral function for particle correlators
	by various authors \cite{Asakawa:2000tr,Aarts:2007pk,Jakovac:2006sf,Rothkopf:2016luz,Ding:2017std,Tripolt:2018xeo,Cyrol:2018xeq,Binosi:2019ecz,Schlichting:2019tbr,PhysRevLett.124.056401}.
	An advantage of the Tikhonov regularization being that it keeps the optimization function a quadratic function, which translates into solving a modified linear system of equations.
	Using a NNLS solver allows solving this system of linear equations while simultaneously enforcing positivity; a feature not shared by other strategies to enforce positivity such as the ansatz $\rho(\omega) = \exp\pqty{\sigma(\omega)}$.
	The most widely used approach is based on the maximum entropy method (MEM) \cite{Asakawa:2000tr}. We will therefore use the same litmus test as in \cite{Asakawa:2000tr} to benchmark our approach, before turning to the actual glueball case.

	\section{The Numerical Method}\label{sec:method}
	
	In a lattice simulation the propagator\footnote{Note the change in notation from $G(p^2)$ to $G(p)$, where $p := p_4$ now stands for the four component of the momentum. See Eq. (\ref{eq:Schwinger}) for the definition of the Schwinger function.} $G(p_n)$
	is computed on a finite set of evenly spaced momenta $p_n$.
	Given a data set $\{ G(p_n) \} := \{ G(p_0), \ldots, G(p_{N-1}) \}$, the Schwinger function is computed using DFFT, resulting in
	a data set $\{ C(\tau_k) \}$, where
	\[ C(\tau_k) = \sum_{n=0}^{N-1} G(p_n) e^{- i 2 \pi k n / N}. \]
	The Laplace transformation to access the spectral function is given by
	\begin{align}\label{eq:schwinger_discrete}
		C(\tau_k) = \mathcal{L}\left\{ \rho(\omega) \right\}(\tau_k) = \int_0^\infty e^{- \omega \tau_k} \rho(\omega) \dd{\omega}.
	\end{align}
	Because $\rho(\omega) \geq 0$ for observable particles, $\pdv*{C}{\tau} \leq 0$. \Cref{eq:schwinger_discrete} can be approximated using the matrix equation
	\begin{equation*}
		\boldsymbol{C} = \boldsymbol{K} \brho,
	\end{equation*}
	with the elements of $\boldsymbol{K}$ defined as
	\begin{equation*}
		\boldsymbol{K}_{k\ell} :=  e^{- \omega_\ell \tau_k} \Delta \omega.
	\end{equation*}
	Since $\brho$ needs to be obtained, and a direct solution is impossible due to the near zero singular values of $\mathbf{K}$, the original
	problem is replaced by the minimisation of the Tikhonov regularizing functional
	\begin{equation}
		J_\alpha = \norm{\boldsymbol{K} \brho - \boldsymbol{C}}_2^2 + \alpha^2 \norm{\brho - \brho^*}_2^2,
		\label{eq:tikhonov}
	\end{equation}
	where $\alpha^2 > 0$ is the Tikhonov parameter and $\brho^*$ is a prior estimate for $\brho$. In order to impose the constraint $\brho \geq 0$ using a NNLS solver, we define
	\begin{align}
		\boldsymbol{A} = \mqty(\boldsymbol{K} \\ \alpha \mathbb{1} ),\quad \boldsymbol{b} = \mqty(\boldsymbol{C} \\ \alpha \brho^*),
	\end{align}
	and rewrite $J_\alpha$ as
	\begin{equation}
		J_\alpha = \norm{\boldsymbol{A} \brho - \boldsymbol{b}}_2^2.
		\label{eq:tikhonov_lsq}
	\end{equation}
	It is straightforward to show that \eqref{eq:tikhonov} and \eqref{eq:tikhonov_lsq} are equivalent. However, the formulation of the problem
	using Eq. \eqref{eq:tikhonov_lsq} can be solved with a non-negative least squares (NNLS) solver such that $\brho \geq 0$ is guaranteed \cite{Lawson1995}.
	In this work we used the implementation of NNLS \cite{Lawson1995} as provided by SciPy \cite{2020SciPy-NMeth}.

We should note that although we work with lattice QCD data, we nevertheless keep using the ``continuum version'' of the spectral relation, as also done in e.g.~\cite{Asakawa:2000tr,Tripolt:2018xeo,Binosi:2019ecz}. The standard MEM paper \cite[Appendix]{Asakawa:2000tr}, compared the usage of an (artificial) lattice kernel against the continuum kernel, and did not find any noticeable difference. This provides justification for this assumption.
	
	\paragraph{Determination of $\alpha$}
	The regularization parameter $\alpha$, in essence, provides a soft threshold to the singular values of $\boldsymbol{K}$, such that the smallest singular values no longer cause numerical issues.
	Choosing $\alpha$ is a delicate affair, since setting it too small means the problem remains ill-defined, whereas setting it too large destroys a lot of the information contained in the data.
	
	Our preferred criterion for $\alpha^2$ relies on the Morozov discrepancy principle \cite{Dudal:2019gvn}, which states that $\alpha^2$ should be chosen such that
	\begin{equation}
		\norm{\boldsymbol{K} \brho - \boldsymbol{C}}_2^2 = \sum_i \sigma_i^2,
		\label{eq:morozov_gl}
	\end{equation}
	where $\sum_i \sigma_i^2$ is the total variance in the data. Intuitively, this criterion implies that the quality of the reconstruction is identical to the quality of the data.
	The $\alpha^2$ obeying \cref{eq:morozov_gl} is guaranteed to be unique \cite{Kirsch:1996:IMT:236740}, and because the value of $\brho$ for a given data set $\boldsymbol{C}$ depends only on $\alpha^2$, any minimization algorithm yields the same solution for $\brho$. We found fast convergence with Nelder-Mead, as implemented in \verb|symfit| \cite{martin_roelfs_2021_5519611}, but we checked that different solvers indeed give numerically the same solution.
	
	\paragraph{Construction of $\boldsymbol{K}$} The matrix $\boldsymbol{K}$ should perform the Laplace transform as truthfully as possible, and therefore $\omega$ should range from $[0, \omega_\text{max})$, where $\omega_\text{max}$ is sufficiently large compared to any features that might appear in $\rho(\omega)$. In order to ensure this we choose to sample $\omega$ evenly in logarithmic space from $[10^{-5}, 10^5]$ \SI{}{\GeV} in $N_\omega$ steps.
	However, the peak positions are very consistent, in both linear and logarithmic space, provided $\omega_\text{max}$ and $N_\omega$ are large enough. 
	Nonetheless, sampling $\omega$ evenly in logarithmic space is preferred, as convergence of the peak positions is reached for much smaller values of $N_\omega$.
	A numerical comparison is provided in \cref{app:numerics}.
	
	\section{Results and Discussion}
	
	In this section we detail the results of applying the method of \cref{sec:method} to various mock and real data.
	In \cref{sec:results_meson} the method is applied to a toy model based on a vector-meson spectral density to establish its reliability. Then,
	in \cref{sec:results_glueball} the method is applied to recent lattice SU(2) propagator data for the $0^{++}$ glueball \cite{Yamanaka:2019yek}.
	
	\subsection{Meson toy-model}\label{sec:results_meson}
	
	\begin{figure}[hbt]
		\centering
		\includegraphics[width=\textwidth,trim={0.5in 0.4in 0.5in 0.5in},clip]{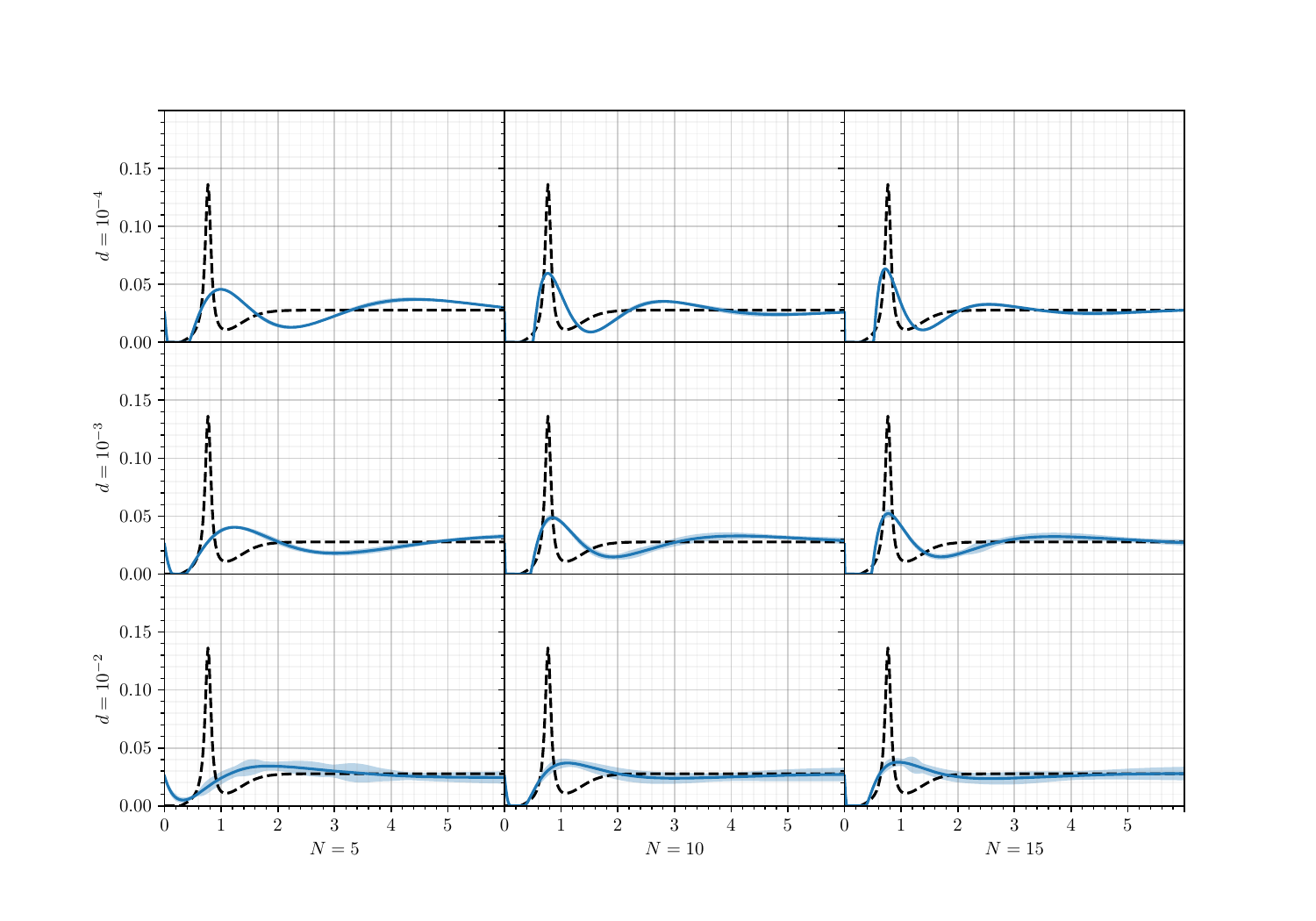}
		\caption{Reconstruction the toy-model spectral density function for various noise levels $d$ as defined in \cref{eq:noise_level} and number of data points $N$. The dashed black line is the original spectral function, while the blue solid curve is the reconstructed spectral function as given by the Tikhonov regularized NNLS method.}
		\label{fig:toymodel}
	\end{figure}
	
	In order to investigate the reliability of the method we consider a realistic toy-model, based on a vector-meson model decay into hadrons,
	as used before in \cite{Asakawa:2000tr}. This particular model needs  a single subtraction ($n=1$), and therefore provides an excellent test of the method.
	To allow comparison of our results to those of \cite{Asakawa:2000tr}, the same process was used to generate the mock data.
	The meson spectral density function is given by
	\begin{equation}
		\rho(\omega) = \frac{2}{\pi}  \bqty{F^2_{\rho} \frac{\Gamma_\rho m_\rho }{(\omega^2 - m_\rho^2)^2 + \Gamma_\rho^2 m_\rho^2} + \frac{1}{8 \pi} \pqty{1 + \frac{\alpha_s}{\pi}} \frac{1}{1 + e^{(\omega_0 - \omega) / \delta}}},
	\end{equation}
	with an energy-dependent width
	\begin{equation}
		\Gamma_\rho(\omega) = \frac{g_{\rho\pi\pi}^2}{48 \pi} m_\rho \pqty{1 - \frac{4 m_\pi^2}{\omega^2}}^{3/2} \theta\pqty{\omega - 2 m_\pi}.
	\end{equation}
	The empirical values of the parameters are
	\begin{alignat*}{3}
		m_\rho &= 0.77 \text{ GeV},\quad & \hspace{1cm} & m_\pi = 0.14 \text{ GeV}, \\
		g_{\rho\pi\pi} &= 5.45, && F_\rho = \frac{m_\rho}{g_{\rho\pi\pi}}, \\
		\omega_0 &= 1.3 \text{ GeV}, && \delta = 0.2 \text{ GeV}.
	\end{alignat*}
    As $\omega \to \infty$, this model behaves like $\rho(\omega \to \infty) = (1 / 4 \pi^2) (1 + \alpha_s/\pi)$. Therefore the integral \cref{eq:propagator_gl} does not converge, and a single subtraction has to be performed; that means $n=1$ in the notation used in \cref{eq:taylor_coeff}.

    Assuming $\alpha_s = 0.3$, the value $(1 / 4 \pi^2) (1 + \alpha_s/\pi) = 0.0277$ can be used as the prior, but identical to \cite{Asakawa:2000tr} we shall use the slightly smaller value $\rho_\text{prior}~=~0.0257$.
    In order to generate mock data, we compute $C_\text{orig}(\tau_k)$ as the Laplace transform of $\rho(\omega)$, on $N$ points $\tau_k$ spaced by $\Delta \tau = 0.085 \text{ fm} = 0.43078 \text{ GeV}^{-1}$.
    The standard deviation of the noise is chosen as
    \begin{equation} \label{eq:noise_level}
        \sigma(\tau_k) = d \, C_\text{orig}(\tau_k) \frac{\tau_k}{\Delta \tau},
    \end{equation}
    where $d$ is a parameter which controls the noise level, identical to that of \cite{Asakawa:2000tr}. The mock data set is then generated as
    \[ C(\tau_k) = \mathcal{N}(\mu=C_\text{orig}(\tau_k), \sigma^2=\sigma\pqty{\tau_k}^2). \]
    These mock data sets were then inverted, ignoring $C(\tau_0)$, for various values of $N$ and $d$, using the method of \cref{sec:method}, to test the robustness of the method. No inversions without positivity constraints were performed, as significant positivity violations were observed in initial trials.  Here we used $N_\omega = 1000$ in the construction of $\boldsymbol{K}$. The results are shown in \cref{fig:toymodel}.
    Both more data-points, or less noise, are found to improve the quality of the reconstruction.
    A direct comparison with \cite[Figure 4]{Asakawa:2000tr} is complicated by the absence of uncertainties on that figure, but the performance of the methods seems comparable to the naked eye.
	
	\subsection{SU(2) Glueball data}\label{sec:results_glueball}
	
	Yamanaka et al.~\cite{Yamanaka:2019yek}, see also \cite{Yamanaka:2019aeq}, have provided us with the Schwinger functions for the SU(2) pure Yang-Mills glueballs, using lattice simulations for $\beta =$ 2.1, 2.2, 2.3, 2.4, 2.5. The Schwinger functions were computed using the raw data, that is, without any smearing applied. The lattice volumes, the number of configurations, the number of Schwinger function time slices $N$, and the uncertainties in each data set, are shown in \cref{tab:configs}. There was no access possible to the configuration per configuration data, implying we have to ignore any possible correlations between the different times. Said otherwise, we work with a diagonal correlation matrix\footnote{N.~Yamanaka, private communication.}.
	
	The simulations of Yamanaka et al.~rely on large ensembles of configurations, resulting in data sets with very low statistical uncertainties. This makes the results of the Tikhonov regularized inversion highly reproducible, as seen in the toy model study of \cref{sec:results_meson}. 
	For further details about the data sets, we refer the reader to \cite{Yamanaka:2019yek}.
	\begin{table}[tb]
		\centering
		\resizebox{\columnwidth}{!}{
			\begin{tabular}{l|l c r c c c} \hline
				$\beta$ & Volume & Volume (\SI{}{\femto\meter^4}) & Configurations & $N$ & $\max(C(\tau)) / \expval{\sigma_C(\tau)}$ & $a \sqrt{\sigma}$ \\
				\hline
				$2.1$ & $10^3 \times 12$ & $(2.73)^3 \times 3.27 = 66.30$ & 1,000,000 & 7 & $3.28 \times 10^{-4}$ & $0.608(16)$  \\
				$2.2$ & $12^4$           & $(2.51)^3 \times 2.51 = 39.87$ & 9,999,990 & 7 & $1.04 \times 10^{-4}$ & $0.467(10)$  \\
				$2.3$ & $14^3 \times 16$ & $(2.31)^3 \times 2.65 = 32.80$ & 4,100,000 & 9 & $1.38 \times 10^{-4}$ & $0.3687(22)$  \\
				$2.4$ & $16^3 \times 24$ & $(1.91)^3 \times 2.86 = 19.90$ & 2,030,000 & 13 & $1.55 \times 10^{-4}$ & $0.2660(21)$  \\
				$2.5$ & $20^3 \times 24$ & $(1.69)^3 \times 2.02 = \,\; 9.72$ & 520,000 & 13 & $3.04 \times 10^{-4}$ & $0.1881(28)$  \\
				\hline
			\end{tabular}
		}
		\caption{The glueball data sets of \cite{Yamanaka:2019yek}. Selection of information taken from \cite[Table~I, Table~II]{Yamanaka:2019yek}. The physical volume in units of \SI{}{\femto\meter^4} was calculated assuming $\sqrt{\sigma} = 0.44$ \SI{}{\giga\electronvolt}.}
		\label{tab:configs}
	\end{table}
	\begin{figure}[tb]
		\centering
		\begin{floatrow}
			\floatbox{figure}[.45\columnwidth][\FBheight][b]
			{\caption{$C(\tau)$ for all data sets. In order to display negative values while keeping log-scale where possible, the range $[-10^{-7}, 10^{-7}]$ is linear.
			Only the data points connected by a dotted line were included in the inversion. 
				}\label{fig:C_all}}
			{\includegraphics[width=0.45\textwidth,trim=0.1in 0.1in 0.1in 0.1in ,clip]{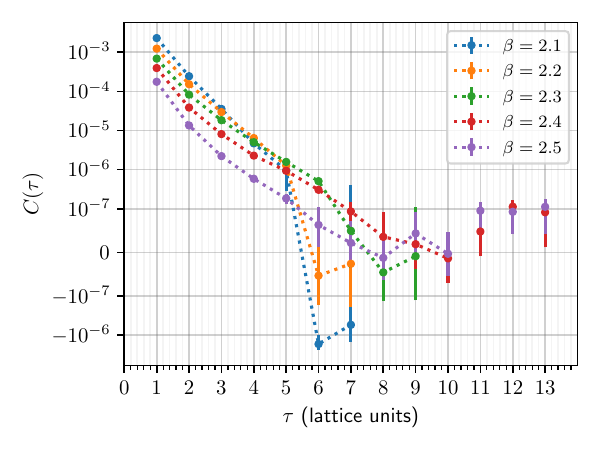}}
			\floatbox{figure}[.45\columnwidth][\FBheight][b]
			{\caption{$\rho(\omega)$ for all data sets, using the $ip$-method. The $\rho(\omega)$ have been normalized to give the last peak an intensity of 1. The colored bands indicate the masses as found by \cite{Yamanaka:2019yek} (see also \cref{tab:glueballmass}).
			The oscillations indicate overfitting and clearly positivity constraints should be imposed.
				}
				\label{fig:rho_all_ip}}
			{\includegraphics[width=0.45\textwidth,trim=0.1in 0.1in 0.1in 0.1in ,clip]{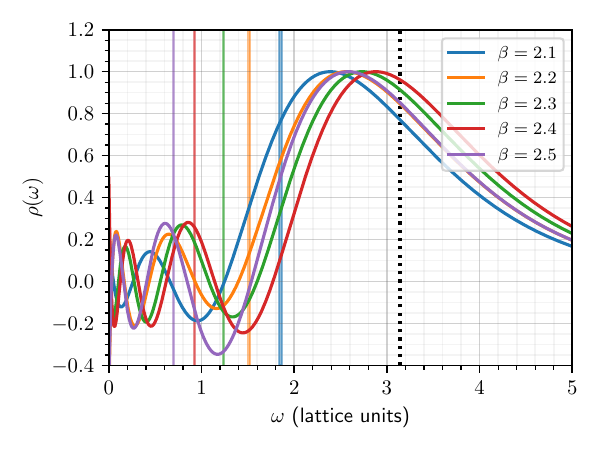}}
		\end{floatrow}
	\end{figure}
	
\begin{figure}[htb]
		\centering
		\begin{floatrow}
			\floatbox{figure}[\FBwidth][\FBheight][c]
			{\caption{$\rho(\omega)$ for all data sets, subject to $\rho(\omega)~\geq~0$, normalized such that the ground state has an intensity of $1$. The colored bands indicate the masses as found by \cite{Yamanaka:2019yek} (see also \cref{tab:glueballmass}). The dotted line at $\omega = \pi$ (lattice units) indicates the largest momentum at which the results can be trusted.}
				\label{fig:rho_all}}
			{\includegraphics[width=0.5\textwidth,trim=0.1in 0.1in 0.1in 0.1in ,clip]{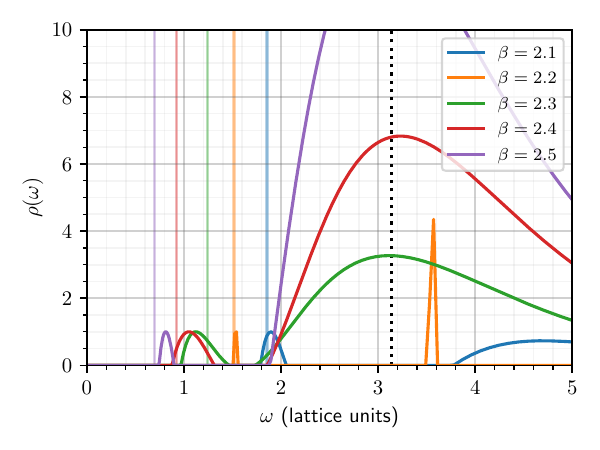}}
			\floatbox{table}[\FBwidth][\FBheight][c]
			{\caption{Maxima of \cref{fig:rho_all} in order of increasing $\omega$, with corresponding left and right Half Width at Half Maximum (HWHM).}\label{tab:states}}
			{
				\small
				\begin{tabular}{|l|l|}
					\hline
					$\beta$ & $\omega \pm \text{HWHM}$ (lattice units) \\
					\hline
					2.1 & $1.895+0.112-0.085$ \\
					& $4.708+1.499-0.747$ \\
					\hline
					2.2 & $1.540+0.018-0.035$ \\
					& $3.571+0.041-0.041$ \\
					\hline
					2.3 & $1.116+0.195-0.110$ \\
					& $3.110+1.653-0.909$ \\
					\hline
					2.4 & $1.053+0.170-0.136$ \\
					& $3.219+1.654-0.914$ \\
					\hline
					2.5 & $0.808+0.068-0.054$ \\
					& $3.039+1.615-0.838$ \\
					\hline
				\end{tabular}
				\vspace{25pt}
			}
		\end{floatrow}
	\end{figure}

	The Schwinger functions for the various simulations can be seen in \cref{fig:C_all}.
	The values of $C(\tau)$ for $\tau > 10$ behave unexpectedly, since $\rho(\omega) \geq 0$ implies $\pdv*{C}{\tau}~\leq~0$. Therefore these data points were excluded from the inversion, as this behavior is likely the result of lattice artifacts.
	On the basis of this argument perhaps even more data points could be excluded, but we found that this does not significantly impact the result, and given the small number of data points available and in the interest of transparency, we have opted to include as many data points as possible in the inversion.
	\Cref{fig:rho_all_ip} shows the spectral density functions as obtained without any constraints on $\rho(\omega)$, using the $ip$-method published in \cite{Dudal:2019gvn}.
	As \cref{fig:rho_all_ip} illustrates, there are significant positivity violations and rapid oscillations in the infrared when positivity is not enforced, especially for small $\omega$, a tell-tale sign of overfitting.\footnote{These should anyhow be taken by a grain of salt given that the finite lattice volume limits the IR resolution.} 
	On the other hand, when positivity is imposed, the resulting spectral functions in \cref{fig:rho_all} all display a clear mass gap that corresponds to a ground state mass in the range of \SI{1.4}{\GeV} to \SI{1.9}{\GeV}. Moreover, the infrared oscillations are gone, and we can actually forget about the deep IR that is anyhow inaccessible to the lattice simulation.
	To produce \cref{fig:rho_all}, we set $N_\omega = 2000$ in the construction of $\boldsymbol{K}$. 
	This is unnecessarily large for reliable extraction of the ground state mass, as convergence is already reached for $N_\omega \geq 1000$ as is shown in \cref{app:numerics}, and was done merely to improve the aesthetics of \cref{fig:rho_all}.
	\Cref{tab:states} lists the $\omega$ values of all the local maxima in \cref{fig:rho_all}, and the left and right Half Width at Half Maximum (HWHM) values of each local maximum.
	
	In order to compare the ground state masses as extracted via the spectral method with the results calculated by Yamanaka et al. \cite{Yamanaka:2019yek}, \cref{tab:glueballmass} lists the two results side by side.
	
	\begin{table}[ht]
		\centering
		\begin{tabular}{l|ll|ll}
			& \multicolumn{2}{c}{Traditional \cite{Yamanaka:2019yek}} & \multicolumn{2}{c}{Spectral representation} \\
			\hline
			$\beta$ & $a m_\phi$  & $m_\phi$ /GeV & $a m_\phi$ &  $m_\phi$ /GeV \\
			\hline
			2.1 & 1.853(13) & 1.341(37) & 1.895(99) & 1.371(80) \\
			2.2 & 1.517(10) & 1.429(32) & 1.540(26) & 1.451(40) \\
			2.3 & 1.241(6) & 1.481(11) & 1.116(153) & 1.331(182) \\
			2.4 & 0.924(8) & 1.528(18) & 1.053(153) & 1.742(254) \\
			2.5 & 0.696(6) & 1.628(28) & 0.808(61) & 1.890(145)
		\end{tabular}
		\caption{The $0^{++}$ glueball ground state masses as presented in Table 3 of \cite{Yamanaka:2019yek} compared with the spectral representation method. The uncertainty in the ground state mass values was calculated by fitting a Gaussian distribution to the entire ground state peak. The physical units were calculated assuming $\sqrt{\sigma} = 0.44$ \SI{}{\GeV}.}
		\label{tab:glueballmass}
	\end{table}

	The uncertainties in the ground state masses as listed in \cref{tab:glueballmass} were calculated by fitting a Gaussian distribution to the ground state peak, and a string tension of $\sqrt{\sigma} = \SI{0.44}{\GeV}$ was assumed to convert to physical units.
	The required values of $a \sqrt{\sigma}$ are given in \cref{tab:configs}, and their uncertainties were compounded with the uncertainties in the ground state mass to calculate the uncertainties in the physical mass $m_\phi$. 
	
	For all but $\beta=2.5$, the scalar glueball masses as obtained by \cite{Yamanaka:2019yek}, and those extracted using the spectral density method, agree within one standard deviation.
	This is noteworthy, since we based ourselves on unsmeared data, whilst the mass estimates from \cite{Yamanaka:2019yek} were obtained after smearing, which usually improves the ground state signal. 
	For $\beta=2.5$ the disagreement is about $\sim 1.8$ standard deviations, which might be explained by the fact that this is also the data set generated from the smallest number of configurations.
    In addition to the ground states, the spectra presented in \cref{fig:rho_all} also hint at first excited states. However, the position of these excited states are close to or beyond $\pi / a$, the largest momentum accessible on the lattice, and thus their estimation has to be treated with care.
	
\begin{figure}[htb]
		\includegraphics[width=0.45\textwidth,trim=0.1in 0.1in 0.1in 0.1in ,clip]{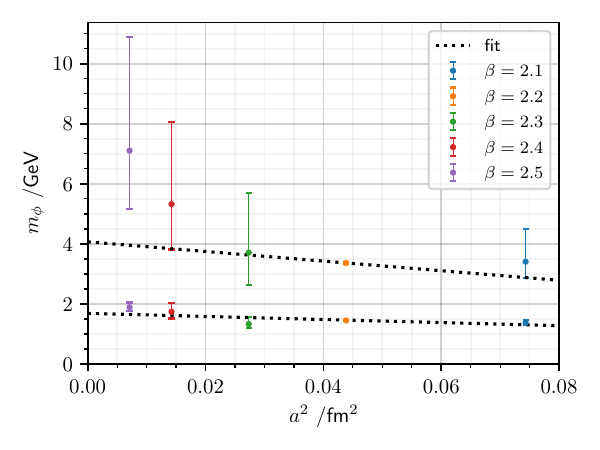}
		\caption{Mass estimates of the ground and excites states, see~\cref{tab:states}, vs $a^2$ in \SI{}{\femto\meter^2}. A weighted linear regression to the data points was performed to estimate the continuum values for the ground and excited state masses.}
		\label{fig:mvsasq}
	\end{figure}

	\Cref{fig:mvsasq} displays the extracted mass estimates as a function of $a^2$, from which an estimate for the continuum values of the masses can be obtained after performing a weighted linear regression to the function $m_\phi(a^2) = s a^2 + m_\phi(0)$. This model is motivated by the fact that the lattice action and operators have corrections of $\order{a^2}$. The continuum estimates for the ground and excited state masses were found to be $m_\phi(0)=1.68 \pm 0.09$ \SI{}{\GeV} and $m_\phi(0)=4.07 \pm 1.01$ \SI{}{\GeV} respectively (where again we assumed $\sqrt{\sigma} = 0.44$ \SI{}{\GeV}).

	The relatively wide peaks for the excited states indicate that we might need a better signal, as there are in principle no decay channels open for these 1st excited states in pure gluodynamics, not even to the $0^{++}$ ground state.
	However, these values are near to or beyond $\omega = \pi / a$, i.e. the resolution of the lattice experiments, and hence are to be taken with a grain of salt.
	Only the $\beta=2.2$ data set displays relatively sharp ground and excited state peaks; we dare to speculate that this is because of the much larger configuration number for $\beta=2.2$ (\cref{tab:configs}). Future research should reveal whether larger configuration numbers and/or smearing techniques result in sharper peaks.
	It is nonetheless reassuring that this crude estimate for the continuum 1st excited state mass comes pretty close to the state of the art prediction of \cite[Table 23]{Athenodorou:2021qvs}, based on independent SU(2) data.
	
	We should also keep in mind that we are aiming at reliable continuum estimates, which implies $\beta$ and the lattice volume should be sufficiently large, so probably values of $\beta \geq 2.4$ are suboptimal from this viewpoint. The size of the (inverse) lattice spacing also determines the maximally accessible momentum scale, forcing our spectral function estimates to vanish in the UV, in contradistinction with the continuum tail going as $\omega^4$, as follows from power counting or explicit computation \cite{Kataev:1981gr}.

	\section{Summary and Conclusion}
	
	In this work we discuss the computation of the K\"all\'en-Lehmann spectral density function $\rho$ from two point correlation functions $G(p^2)$
	that can have a (divergent) polynomial part. The extraction of the finite part of $G(p^2)$ that is associated with $\rho$ requires the subtraction
	of that polynomial part. To avoid doing the subtraction directly on the numerical data, an operation that can be cumbersome and a source
	of large uncertainties, we consider the Schwinger function $C(\tau)$ instead of $G(p^2)$.
	Once the Schwinger function has been calculated from $G(p^2)$, subtractions are performed simply by removing the data point $C(0)$.
	The spectral density is then obtained by taking the inverse Laplace transform of $C(\tau)$ for $\tau > 0$. 
	For this inverse Laplace transform, the inversion method as described in \cite{Dudal:2019gvn}, that does not constrain $\rho$, was compared with a reformulation of the method
	that relies on a non-negative least square solver to impose a positive $\rho$.
	
	The numerical method was first tested on mock data from a given spectral density. The test followed the same rules as used
	in the original maximum entropy method inversion paper \cite{Asakawa:2000tr}, with similar results.
	
	The inversion method was then applied to extract the K\"all\'en-Lehmann spectral density function from the
	lattice SU(2) glueball propagator data with quantum numbers $J^{PC} = 0^{++}$, based on the unsmeared data of \cite{Yamanaka:2019yek}. We found that the inversion method is robust
	and allows estimation of the ground state and provides hints of excited state masses. The comparison of the mass estimates
	between a conventional lattice approach \cite{Yamanaka:2019yek} and the method described herein shows that the latter achieves
	comparable results for the ground state, despite the lack of smearing. 
	
	The new spectral method is sufficiently general to be applied to any spectroscopic calculation. In future research, we will scrutinize the usage of smeared data to boost the contribution of the lowest lying states, which should allow access to at least the 1st excited state with higher accuracy. Afterwards, the analysis performed in the current manuscript might also be extended to the analysis of the spectra of other glueball states with different quantum numbers.
	
	\section{Acknowledgements}
	The authors are indebted to N.~Yamanaka, H.~Iida, A.~Nakamura and M.~Wakayama for sharing the Schwinger function estimates based on their data of \cite{Yamanaka:2019yek}. We also thank A.~Athenodorou and M. Teper for useful communication.
	
	O.O.~was partly supported by the FCT – Funda\c{c}\~ao para a Ci\^encia e a Tecnologia, I.P., under project numbers UIDB/04564/2020 and UIDP/04564/2020, while the work of D.D.~and M.R.~was supported by KU Leuven IF project C14/16/067.

	\newpage
	\appendix
	\section{Subtracted correlator}\label{app:subtractions}
	A correlator $G(p^2)$ can be expressed in terms of a K\"all\'en–Lehmann spectral representation as
	\begin{equation}
		G(p^2) = \int_{0}^{\infty} \frac{2 \omega \rho(\omega) \dd{\omega}}{\omega^2 + p^2},
	\end{equation}
	if this integral converges. However, if the correlator contains divergences polynomial in $p^2$,  it is still possible to find a K\"all\'en–Lehmann spectral representation, if one first subtracts the polynomial part. In order to do this, consider the Taylor expansion of $G(p^2)$ around $\bar{p}$:
	\begin{align}
		G(p^2) = \sum_{k=0}^{n-1} (p^2-\bar{p}^2)^{k} \pdv[k]{G(p^2)}{(p^2)}\Bigg\vert_{p^2=\bar{p}^2} + \tilde{G}_n(p^2).
	\end{align}
	where $\tilde{G}_n(p^2)$ is the remainder after the first $n$ terms have been isolated.
	At zeroth order, we find
	\begin{align}
		\tilde{G}_1(p^2) &= G(p^2) - G(\bar{p}^2) \notag \\
		&= (- p^2  + \bar{p}^2) \int_{0}^{\infty} \frac{2 \omega \rho(\omega) \dd{\omega}}{(\omega^2 + \bar{p}^2)(\omega^2 + p^2)}
	\end{align}
	Through induction, we obtain
	\begin{equation}
		\tilde{G}_n(p^2) = (- p^2 + \bar{p}^2)^n \int_{0}^{\infty} \frac{2 \omega \rho(\omega) \dd{\omega}}{(\omega^2 + \bar{p}^2)^n(\omega^2 + p^2)}
	\end{equation}
	This allows us to state that
	\begin{align}
		G(p^2)
		&= \sum_{k=0}^{n-1} a_{k} (p^2-\bar{p}^2)^{k} +  (- p^2 + \bar{p}^2)^{n} \int_{0}^{\infty} \frac{2 \omega \tilde{\rho}(\omega) \dd{\omega}}{\omega^2 + p^2} \notag \\
		&=  \sum_{k=0}^{n-1} a_{k} (p^2-\bar{p}^2)^{k} +  (- p^2 + \bar{p}^2)^{n} \int_{-\infty}^{\infty} \frac{\tilde{\rho}(\omega) \dd{\omega}}{\omega - ip}
	\end{align}
	with
	\begin{align}
		a_{n} &= \frac{1}{n!} \pdv[n]{G(p^2)}{(p^2)}\Bigg\vert_{p^2=\bar{p}^2}, \\ 
		\tilde{\rho}(\omega) &= \frac{\rho(\omega)}{(\omega^2 + \bar{p}^2)^n}.
	\end{align}
	Because $\rho$ is an odd (real) function for bosonic degrees of freedom \cite{Asakawa:2000tr}, so is $\tilde{\rho}$.
	We will now calculate the Fourier transform of the subtracted correlator, in order to obtain $C(\tau)$.
	\begin{align}
		\F{G(p^2)} &=  \underbrace{\sum_{k=0}^{n-1} a_{k} \F{(p^2)^{k}}}_{A} +  \underbrace{\F{(- p^2)^{n} \int_{-\infty}^{\infty} \frac{\tilde{\rho}(\omega) \dd{\omega}}{\omega - ip}}}_{B}
	\end{align}
	where we have set $\bar{p}=0$. For the first term $A$ we obtain
	\begin{align}
		A = \sum_{k=0}^{n-1} a_{k} \F{(p^2)^{k}} &=  2 \pi \sum_{k=0}^{n-1} a_{k} (-1)^k \delta^{(2k)}(\tau)
	\end{align}
	and for the second
	\begin{align}
		B &= \F{(- p^2)^{n} \int_{-\infty}^{\infty} \frac{\tilde{\rho}(\omega) \dd{\omega}}{\omega - ip}} \\
		&= \int_{0}^{\infty} \dd{\omega} \tilde{\rho}(\omega) \F{  \frac{2 \omega (- p^2)^{n}}{\omega^2 + p^2}} \\
		&= 2 \pi (-1)^{n} \int_{0}^{\infty} \dd{\omega} \tilde{\rho}(\omega) \left[ -2 \sum_{k=2}^{n} \omega^{2k-3} \delta^{(2(n-k)+2)}(\tau) + \omega^{2n} e^{- |\tau| \omega} \right] \notag \\
		&= 4 \pi (-1)^{n+1} \sum_{k=2}^{n} \delta^{(2(n-k)+2)}(\tau) \int_{0}^{\infty} \dd{\omega} \omega^{2k-3} \tilde{\rho}(\omega) + \mathcal{L}\left\{ \tilde{\rho}(\omega) \omega^{2n} \right\}(|\tau|)  \notag
	\end{align}
	As a result the total transform is given by
	\begin{align}
		\F{G(p^2)} &= 2 \pi \sum_{k=0}^{n-1} a_{k} (-1)^k \delta^{(2k)}(\tau) + \mathcal{L}\left\{ \rho(\omega) \right\}(|\tau|) \\
		&\quad + 4 \pi (-1)^{n+1} \sum_{k=2}^{n} \delta^{(2(n-k)+2)}(\tau) \int_{0}^{\infty} \dd{\omega} \omega^{2k-3} \tilde{\rho}(\omega) \notag
	\end{align}
	An important consequence of this is that all the subtractions are turned into (derivatives of) delta functions, which end up at $\tau=0$. Therefore, taking only $\tau>0$, we can say that
	\begin{align}
		C(\tau) = \mathcal{L}\left\{ \rho(\omega) \right\}(\tau).  \notag
	\end{align}
	
	\newpage
	\section{Numerical dependence on \texorpdfstring{$N_\omega$}{Nomega}}\label{app:numerics}
	
	To produce the glueball spectra given in \cref{sec:results_glueball}, $\omega$ was sampled evenly in logarithmic space between $[10^{-5}, 10^5]$ \SI{}{\GeV}, in $N_\omega=2000$ steps, when constructing the Laplace kernel $\boldsymbol{K}$. However, this number of steps $N_\omega$ is unnecessarily large, and was chosen merely to generate aesthetically pleasing graphs.
	
\begin{figure}[htb]
		\centering
		\begin{subfigure}[b]{0.48\textwidth}
			\includegraphics[width=\textwidth,trim=0 0.15in 0 0.1in ,clip]{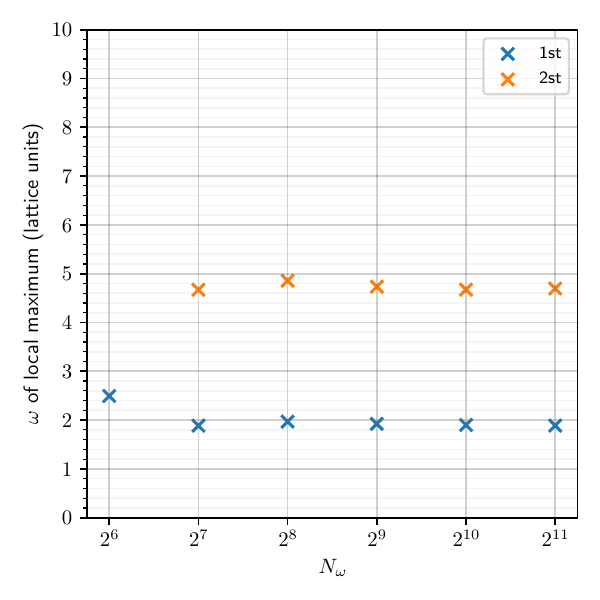}
			\caption{Results for logarithmic sample of $\omega$ in $[10^{-5}, 10^5]$ \SI{}{\GeV}.}
			\label{fig:nsteps_log}
		\end{subfigure}
		\begin{subfigure}[b]{0.48\textwidth}
			\includegraphics[width=\textwidth,trim=0 0.15in 0 0.1in ,clip]{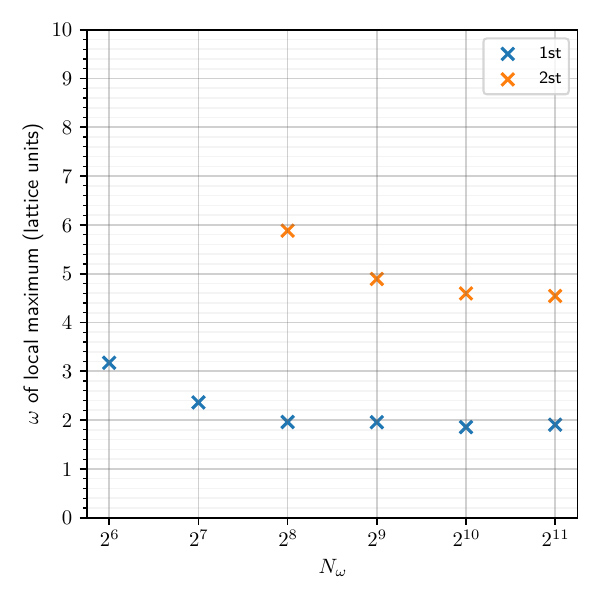}
			\caption{Results for linear sample of $\omega$ in $[0, 100]$ \SI{}{\GeV}.}
			\label{fig:nsteps_lin}
		\end{subfigure}
		\caption{The position of the first and second maxima for the $\beta = 2.1$ data set, as a function of the number of samples $N_\omega$ in the Laplace kernel $\boldsymbol{K}$.}
		\label{fig:nsteps}
	\end{figure}

\begin{figure}[htb]
		\centering
		\begin{subfigure}[b]{0.49\textwidth}
			\includegraphics[width=\textwidth,trim=0 0.1in 0 0.1in ,clip]{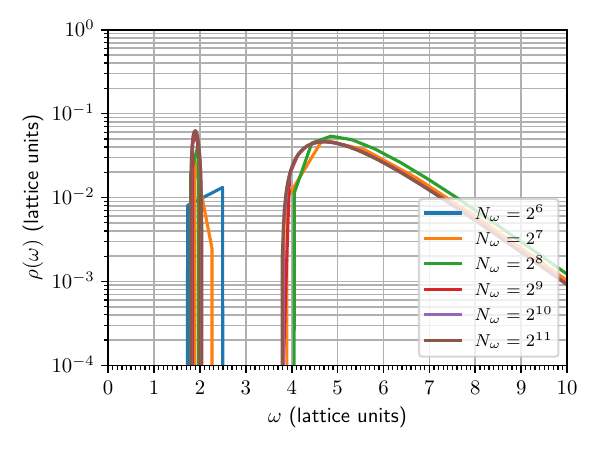}
			\caption{$\omega$ sampled uniformly in logarithmic space between $[10^{-5}, 10^5]$  \SI{}{\GeV}.}
			\label{fig:rho_nsteps_log}
		\end{subfigure}
		\begin{subfigure}[b]{0.49\textwidth}
			\includegraphics[width=\textwidth,trim=0 0.1in 0 0.1in ,clip]{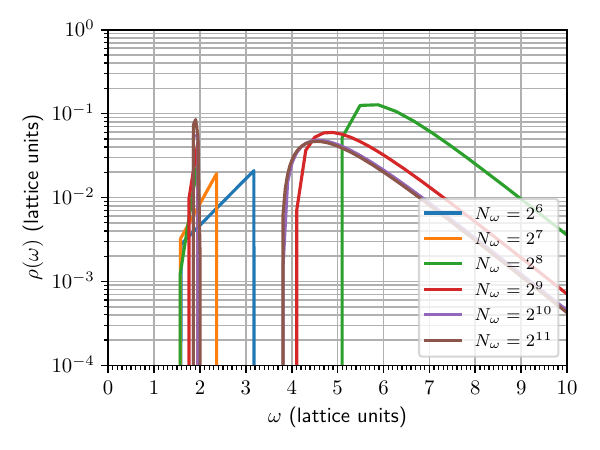}
			\caption{$\omega$ sampled uniformly in linear space between $[0, 100]$  \SI{}{\GeV}.}
			\label{fig:rho_nsteps_lin}
		\end{subfigure}
		\caption{$\rho(\omega)$ for the $\beta=2.1$ data set, as a function of the number of samples $N_\omega$ in the Laplace kernel $\boldsymbol{K}$.}
		\label{fig:rho_nsteps}
	\end{figure}

	To demonstrate the convergence of the maxima of the spectral functions, \cref{fig:nsteps} shows the $\omega$ coordinate of the first and second maximum in the $\beta=2.1$ data set for increasing $N_\omega$. In both the linearly spaced and logarithmically evenly spaced scenarios, the ground state converges rapidly, and the second maximum soon follows suit.
	In addition, \cref{fig:rho_nsteps} shows the full spectrum in both scenarios.
	There is excellent agreement between the two methods for the determination of the ground state mass, but the overall convergence is better when using logarithmically evenly spaced samples.
	

\printbibliography

\end{document}